\documentclass[article,twocolumn,english,secnumarabic,footinbib,tightenlines,nobibnotes,aps,prb,unsortedaddress,showpacs,reprint]{revtex4}
\usepackage{graphicx}
\usepackage{amsmath,amssymb}
\DeclareMathOperator{\Tr}{Tr}
\usepackage{color}
\usepackage{float}
\usepackage{hyperref}

\definecolor{Zcolour}{rgb}{0.992, 0.588, 0.22}

\def\cO{{\cal O}}
\def\cA{{\cal A}}
\def\cH{{\cal H}}
\def\cL{{\cal L}}

\def\Tr{{\rm Tr}}
\def\Im{{\rm Im}}
\def\be{\begin{equation}}
\def\ee{\end{equation}}
\def\bea{\begin{eqnarray}}
\def\eea{\end{eqnarray}}

\begin{document}

\title{Spin transport of weakly disordered Heisenberg chain at infinite temperature}
\author{Ilia Khait}
\affiliation{Physics Department, Technion, 32000 Haifa, Israel}
\author{Snir Gazit}
\author{Norman Y. Yao}
\affiliation{Department of Physics, University of California, Berkeley, CA 94720, USA}
\author{Assa Auerbach}
\affiliation{Physics Department, Technion, 32000 Haifa, Israel}
\date{\today}

\begin{abstract}
We study the disordered Heisenberg 
spin chain, which exhibits many body localization at strong disorder, in the weak to moderate disorder  regime.   
A continued fraction calculation of dynamical correlations
is devised, using a variational extrapolation of recurrents. Good convergence for the infinite chain limit is shown.
We find that the local spin  correlations decay at long times as $C \sim t^{-\beta}$, while the conductivity 
exhibits a low frequency power law $\sigma \sim  \omega^{\alpha}$. The exponents depict sub-diffusive behavior $ \beta < 1/2, \alpha> 0 $ at {\em all finite disorders},
and convergence to the scaling result, $\alpha+2\beta = 1$, at large disorders.  
\end{abstract}
\maketitle

\section{Introduction}

Single band disordered electrons in one dimension are localized at all temperatures~\cite{anderson_loc}. 
In the presence of interactions~\cite{MBLReview}, recent progress has shown the existence of a many-body localized (MBL) phase  at strong disorders, even for high temperatures, while a transition to a delocalized phase occurs as the disorder is weakened~\cite{BaskoMBL, OganesyanHighT,ZnidariED,Luitz2015,RajivNLC}.
This MBL phase is marked by a slow logarithmic growth of entanglement entropy after a quench~\cite{BardarsonLogEnt, SerbynSlow,ZnidariED}, and the emergence of local integrals of motion~\cite{SerbynLocalIntegrals, HusePheno,VoskIM,imbrie2014many}. 
Certain features of many-body localization have already been experimentally observed in cold atomic systems \cite{SchreiberMBLColdAtoms,DeMarco} and trapped ions \cite{MBLTrappedIons}, while a theoretical  renormalization group analysis \cite{VoskRSRG,PotterRSRG} predicts a continuous MBL transition characterized by a diverging dynamical critical exponent.

At finite but weak disorder, a delocalized thermal phase that exhibits sub-diffusive transport was found~\cite{PalHuseMBL,AletL22,AgarwalSubDiff,BarLevSubDiff,Luitz2015,EnergyDiffusion,Prelovsek}.
Specifically, the local spin excess decays in time as $t^{-\beta}$, where  $\beta < 1/2$ and was seen to vanish continuously at the MBL transition. 
To account for this sub-diffusive behavior, a Griffiths mechanism was proposed~\cite{AgarwalSubDiff}, where the long time dynamics are dominated by the existence of rare but large insulating regions. 	
In the clean limit, at high temperature, spin transport is believed to have a diffusive component~\cite{affleckPRL,AffleckHighTDiff,BhomMuller}, i.e. $\beta=1/2$, with a small or possibly vanishing Drude weight~\cite{JoelReview,SteinigewegPRB}.
An important question remains: {\em Is  there  a finite diffusive interval in the weak disorder regime?}   

The long-time response of the delocalized phase is difficult to access with current numerical tools.
Exact Diagonalization (ED) is limited to small chains of order 20 sites, and  Density Matrix Renormalization Group (DMRG) requires a short entanglement length, which is characteristic of the MBL phase
but not the delocalized regime. Therefore, answering this question requires a different approach. 

\begin{figure}[b!]
	\includegraphics[width=0.79\columnwidth]{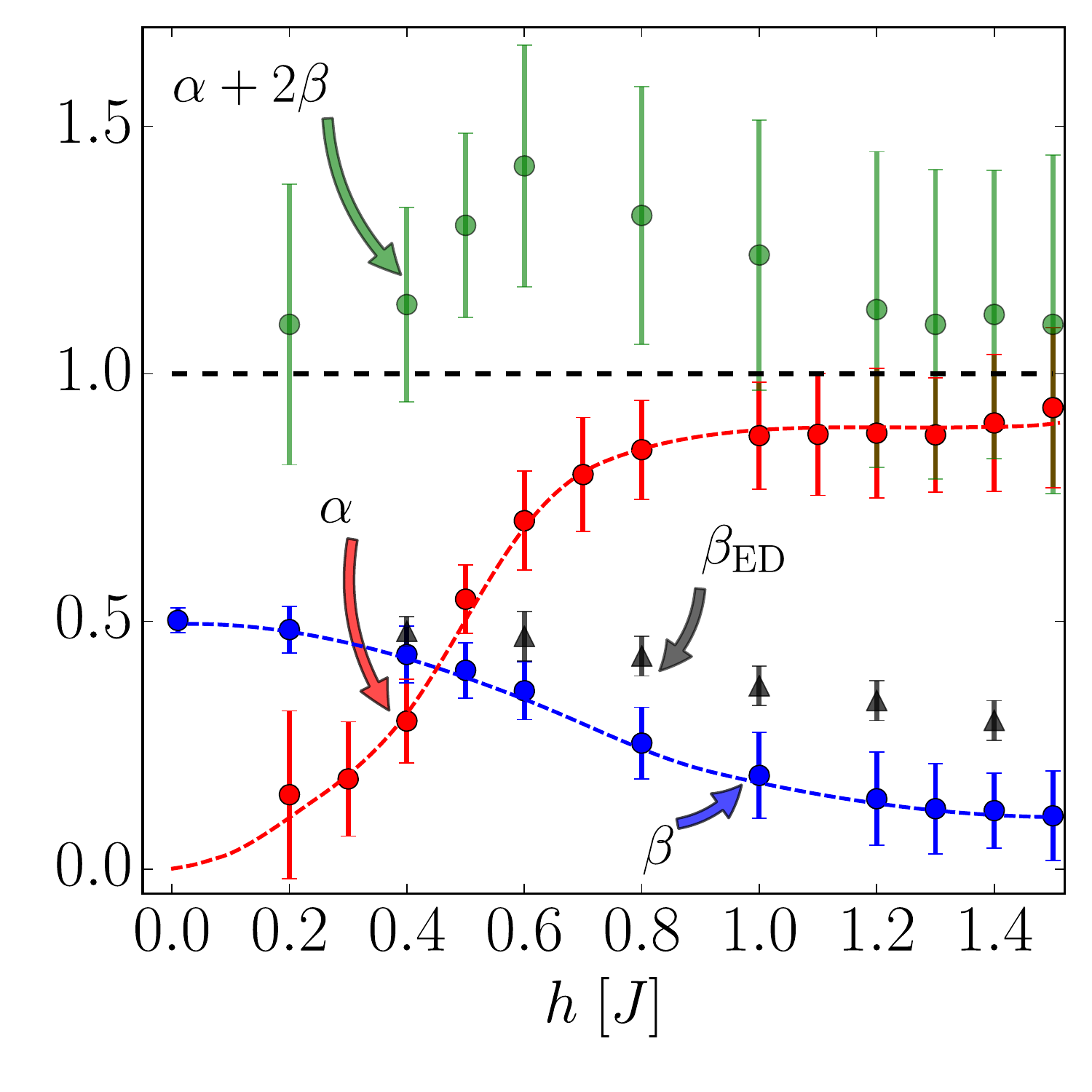}
	\caption{(Color online) Average exponents for  disordered Heisenberg model, evaluated by the continued fractions VER method. $h$ is the disorder strength.  
	The local spin correlations decay  in time as $t^{-\beta}$,  ($\beta$ in blue circles). The low frequency conductivity rises as $\omega^\alpha$ ($\alpha$ in red circles).
Comparison is made with exact diagonalization results on 22 sites $\beta_{\rm ED}$ (black triangles). Error bars are given by least square fit (see Subsection~\ref{subsec:errorest}). The sum  $\alpha+2\beta$ converges to unity at higher disorder - as expected by  scaling (Eq.~\eqref{eq:scaling}).  
\label{fig:PhaseDiagram}}
\end{figure}

In this article, we calculate the infinite temperature dynamical correlations, using a newly developed method: Variational Extrapolation of Recurrents (VER).
Our approach uses the continued fraction representation. A large, but finite, set of recurrents is computed by tracing over commutators of the Hamiltonian with the relevant spin operators.
The remaining recurrents require an extrapolation scheme. Here, we extend the commonly used Gaussian termination approximation~\cite{viswanath_recursion_book,lindnerHCB}, 
to a  family of variational functions chosen to satisfy general physical considerations (i.e. positivity, high frequency decay, etc.)
The accuracy of VER functions and their convergence with the number of computed recurrents, is tested and discussed. 

\begin{figure}[h]
	\includegraphics[width=\linewidth]{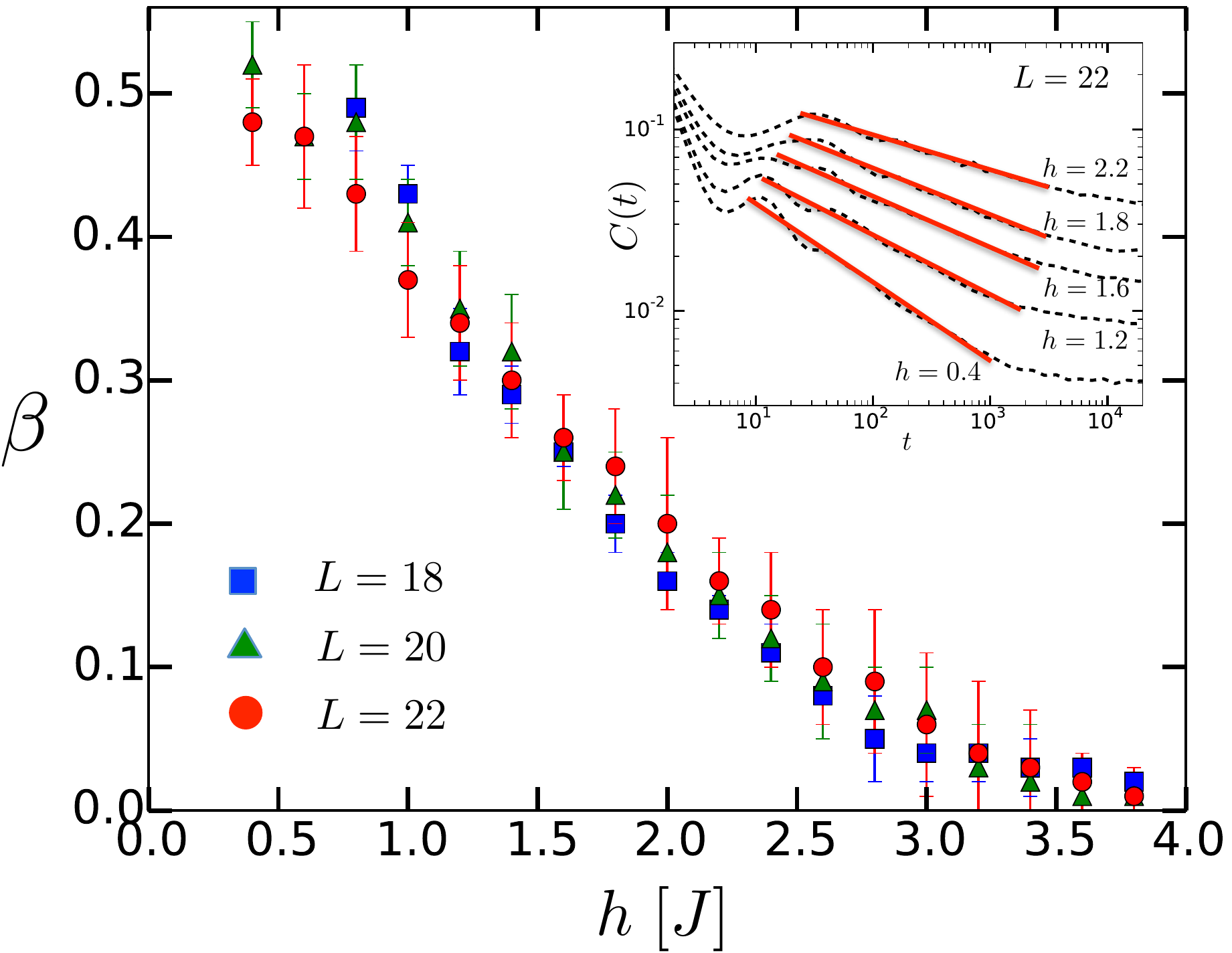}
	\caption{(Color online) The local spin excess decay exponent, $\beta$, of the disordered Heisenberg model as extracted from ED. (Blue squares) represent system size $N=18$, (green triangles) $N=20$, and (red circles) $N=22$. One observes a sub-diffusive exponent well before the MBL transition, which occurs in the data at $h/J\sim3.5$. Despite the relatively large system sizes, it is difficult to extract exponents for $h/J < 0.4$. We utilize $1000$ disorder realizations for $N=18$, $300$ for $N=20$ and $44$ for $N=22$. 
		The inset shows the extraction of $\beta$ for $N=22$ at various disorder strengths. 
		\label{fig:L22ED}}
\end{figure} 

We compute the local dynamical spin correlation function and the ac-conductivity of the one dimensional random field Heisenberg model. 
Our main results are:
(i) In the clean limit, the local spin correlations decay  with $\beta=0.541 \pm 0.065 $,  which confirms the expected diffusive behavior.
Our method achieves a much higher accuracy than previous estimates of $\beta=0.37 \pm 0.12$~\cite{BhomMuller, Fabricius1998}.    
(ii) At finite disorder, the spin transport is {\em sub-diffusive throughout the delocalized regime} (See Figures~\ref{fig:PhaseDiagram},\ref{fig:L22ED}). The ac-conductivity exponent ($\sigma(\omega) \sim \omega^\alpha$) is consistent with the scaling relation 
$\alpha + 2\beta =1$~\cite{AgarwalSubDiff}. Thus, we conclude that there is no diffusive phase at any finite disorder. 

The paper is organized as follows. In Sec.~\ref{sec:model}, we introduce the random field Heisenberg model and the response functions we study. Section~\ref{sec:method} introduces the continued fraction representation of the correlation functions and explains the VER algorithm under generic settings. Section~\ref{sec:results} describes the results obtained by the VER method and provides a comparison with an extensive ED study.  In addition, we discuss the error estimation of the extrapolation scheme. Finally, we  conclude in Sec.~\ref{sec:discusion}, where we discuss the strengths and limitations of the method.
	
\section{\label{sec:model} Model and Observables}
We study the dynamical response of the random field Heisenberg model, which serves as a minimal model
of disordered and interacting fermions or hard core bosons in one dimension~\cite{ColemanBook},
\be
\cH=J \sum_{i=1}^N \vec{S}_i\cdot \vec{S}_{i+1} +\sum_i h_i S_i ^z
\label{eq:Ham}
\ee
with $\vec{S}_i$ a spin-half operator.
The local magnetic fields $h_i$ are independent and distributed uniformly $h_i \in [-h ,h]$.
The infinite temperature local autocorrelation function is
\be
C (t)= 2^{-N} \text{Tr} \left(  S_{N/2}^z(t)S_{N/2}^z(0)\right)    \sim t^{-\beta}  
\label{eq:Ci}
\ee
where $N$ is the number of sites, the time dependence of an operator is due to the Heisenberg picture, and the dynamical  conductivity  at  high temperature $T$  in the Lehmann representation is
\be
T\sigma(\omega) = \pi 2^{-N} N \sum_{n\ne m} |\langle n| I |m\rangle|^2 \delta(E_n-E_m-\omega)
\label{eq:TS}
\ee
where $n,m$ are the eigenstates of the system and $E_n,E_m$ the corresponding eigenenergies. The spin current operator is
\be
I \equiv   4 J N^{-1} \sum_{i=1}^N (S^x_i S^y_{i+1} -S^y_i S^x_{i+1} ).
\label{eq:Current}
\ee

If we add and subtract the $n=m$ term in Eq.~\eqref{eq:TS}, we can write the conductivity as 
\bea
T\sigma(\omega) &=& N \Im \left\lbrace  2^{-N} \Tr \left(  I {1\over \omega - \cL +i0^+} I \right) \right\rbrace  \nonumber \\ &-& \pi N 2^{-N}  \delta(\omega) \sum_{n} |\langle n| I |n\rangle|^2,
\label{eq:TS1}
\eea
where  $\cL$ is the Liouvillian, which acts on an operator as $ \cL \cA = [\cH,\cA]$. 

Note that $\sigma(\omega) $ is a long wavelength response of the system, while $C(t) $ is highly local. 
At the  MBL transition~\cite{AgarwalSubDiff,BarLevSubDiff,VoskRSRG,PotterRSRG}, it is expected that diffusion will be arrested and $\beta \to 0$. 
According to Ref.~\cite{AgarwalSubDiff}, in the delocalized regime, if  space and time  scales are simply related, then the structure factor 
will obey $C(q,\omega) \sim \omega^{-1} g(q/\omega^\beta)$ where $ g(q/\omega^\beta)$ is a universal scaling function. 
This implies that  the dynamical critical exponent is $z=1/\beta$.
By the continuity equation, the $q$-dependent conductivity obeys $\sigma(q,\omega) \sim \omega^2 \partial_q^2 C(q,\omega) = \omega \partial_q^2 g(q/\omega^\beta)$.
Thus, it follows that $\sigma(q=0,\omega) \sim \omega^{1-2\beta}$, which results in the scaling relation 
\be
\alpha+2\beta=1.
\label{eq:scaling}
\ee 
Here, we compute $\alpha$ and $\beta$ as a function of the disorder strength $h/J$. 

\section{\label{sec:method} Numerical Method}
\subsection{\label{subsec:cont-frac} Continued Fraction Representation}

The autocorrelations~\cite{comment1}~\nocite{Perk,McCoy} of any operator $\cO$ at $T\to \infty$ are described by
\be
C(t) =   2^{-N} \Tr  \langle \cO(t) \cO(0) \rangle 
\ee
The imaginary part of the Fourier transform of $C(t)$, $C''(\omega)$, defines a set of moments
\be
\mu_{2k}=\frac{1}{2\pi}\int_{-\infty}^{\infty}d\omega \omega^{2k} C''(\omega) ,~~~k=0,1,2\ldots \infty 
\label{eq:moments}
\ee
which can be computed at infinite temperature as traces of operators
\be
\mu_{2k} =  2^{-N} \text{Tr} \left(  \cO^\dagger \cL^{2k} \cO \right)
\label{eq:mu2k}
\ee
These moments (Eq.~\eqref{eq:moments}) are the Taylor expansion coefficients of $C(t)$.
Quite surprisingly, the same moments  also encode information about {\em long timescales}, i.e.  low frequency fluctuations.

The continued fraction representation of the complex correlation function \cite{Mori1965,Lee1982,viswanath_recursion_book}, is
\bea
\label{eq:contfrac}
C(z) &=&   2^{-N} \Tr \left(  \cO^\dagger  {1\over z - \cL } \cO  \right)\nonumber\\
&=& \cfrac{2 \mu_0}{z-\cfrac{\Delta_1^2}{z-\cfrac{\Delta_2^2}{z-\ddots}}}
\eea  
Setting $z\to \omega+i0^+$ defines 
\be
C(z=\omega+i0^+) = C'(\omega) -i C''(\omega) 
\ee
where $C', C''$  are the real and imaginary parts of $C(z)$ and are related by a Hilbert (Kramers-Kronig) transform.
Any finite set of  {\em recurrents}  $\{ \Delta_1, \Delta_2, \ldots \Delta_{n_{\rm max}} \}$  is {\em algebraically determined}  by the 
same number of moments $\{ \mu_2, \mu_4,\ldots  \mu_{2 n_{\rm max}} \}$.
It is easy to see that the local spin correlation function (which is the Fourier transform  of Eq.~\eqref{eq:Ci}), and the finite frequency conductivity,  Eq.~\eqref{eq:TS1}, can both be expressed as continued fractions. 

In the following, we consider a lattice with periodic boundary conditions of length $N>n_{\rm max}$. This ensures that after applying $\cL^{n_{\rm max}}$ on $\mathcal{O}$, none of the generated operators encircle the chain, hence finite size effects are avoided.

\subsection{Recurrents Calculation}

To reconstruct the response function, $C(\omega)$, using a continued fraction (Eq.\eqref{eq:contfrac}) approach, we must calculate the recurrents $\Delta_n$. This is achieved by a Gram-Schmidt procedure in the Operator Hilbert Space (OHS).

The OHS is spanned by spin-half operators of the form $S_{i_1}^{\alpha_{i_1}} S_{i_2}^{\alpha_{i_2}} \ldots S_{i_k}^{\alpha_{i_k}}$. Where $i_k$ is a lattice site index and $\alpha_i=x,y,z$. The infinite temperature inner product between two operators $A,B$ belonging to the OHS is defined as 
\be
\left( A, B \right) = 2^{-N} \Tr \left [  A^\dagger B  \right ] 
\ee
where $N$ is the number of sites. Following these definitions we recursively construct a set of orthogonal operators $\left\lbrace \hat{O}_i \right\rbrace $ using a Gram-Schmidt orthogonalization
\bea 
\hat{O}_{n+1} & = & c_{n+1}\left(\mathcal{L}\hat{O}_{n}-\Delta_{n}\hat{O}_{n-1}\right) \nonumber \\
\Delta_{n} & = & \left(\hat{O}_{n},\mathcal{L}\hat{O}_{n-1}\right) \nonumber \\
c_{n+1} & = & \left[\left(\hat{O}_{n},\mathcal{L}^{2}\hat{O}_{n}\right)-\left|\Delta_{n}\right|^{2}\right]^{-1/2}
\label{eq:GS}
\eea
with $\Delta_0=0$ and $c_{n}$ is a normalization factor. $\mathcal{L}$ is the Liouvillian defined by $\mathcal{L}A= \left[ \mathcal{H}, A \right] $, square brackets stand for a commutator.

Returning to our case, for the local spin correlations $\hat{O}_0 = S^z_i$, and for the conductivity $\hat{O}_0 = {(I,I)^{-1/2}}~I$ with the spin current operator defined in Eq.~\eqref{eq:Current}.

\begin{figure}[ht]
   \includegraphics[height=2.3in]{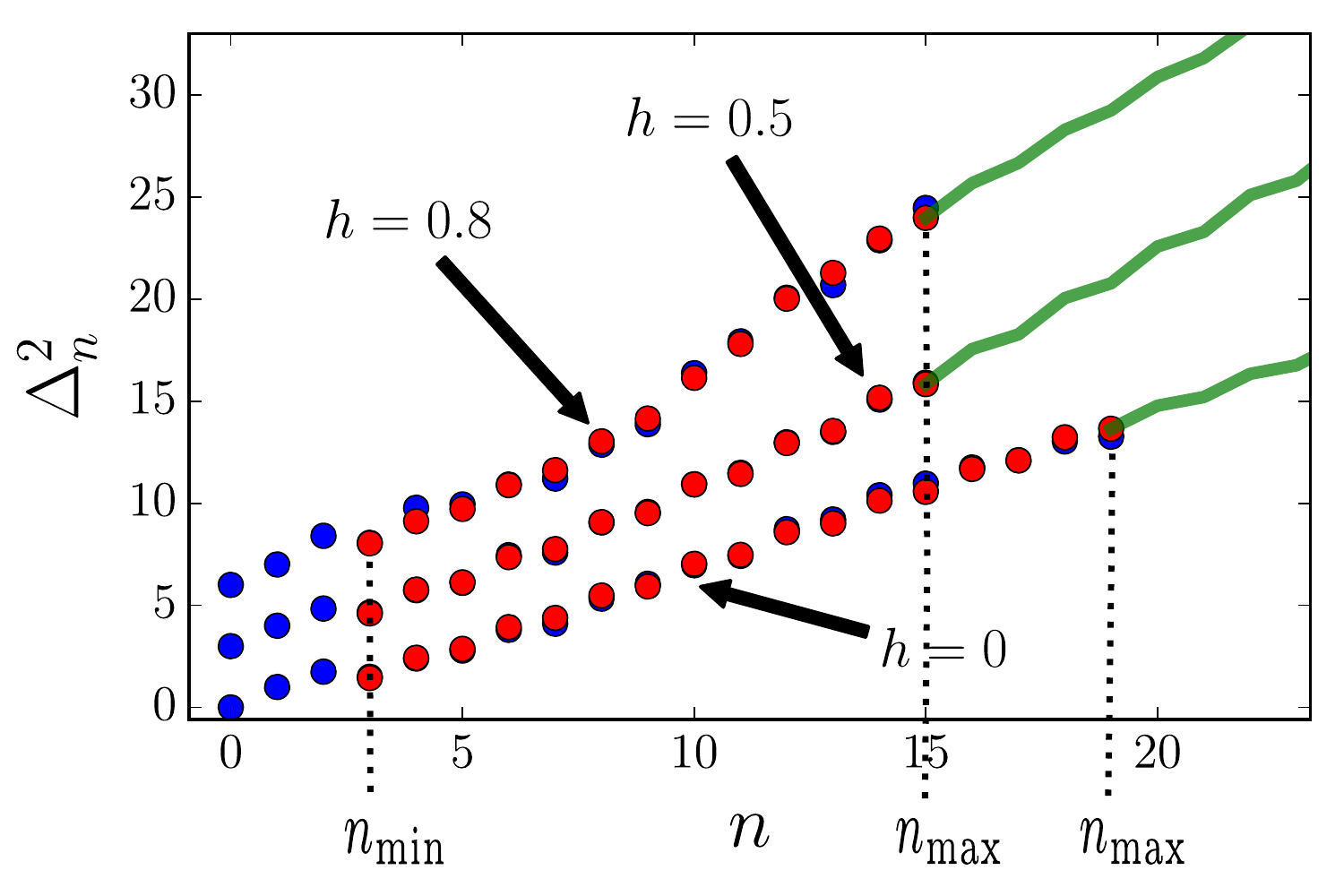}  
   \caption{(Color online) The VER calculation of the local spin correlations.  Blue circles are the calculated recurrents $\Delta_n^2$ for a particular disorder realization of strength $h$. Red circles are  variational recurrents $\tilde{\Delta}^2_n$.
 Green solid lines are the extrapolated higher order recurrents within the VER scheme (see text). The vertical displacement is artificial, where $\Delta_0=0$. Even -  odd alternation of the recurrents reflect the asymptotic {\bf low} frequency power law of the correlation function.
  \label{fig:RecExtrapolation}}  
\end{figure} 

\subsection{\label{subsec:VER}Variational Extrapolation of Recurrents (VER) Procedure}

To evaluate Eq.~\eqref{eq:contfrac}, for {\em any} frequency $\omega$, the full  (infinite) set of moments or recurrents is needed. In practice, only a finite number of recurrents can be computed using this procedure since repeatedly applying the Liouvillian leads to a factorial growth in the number of operators in $\hat{O}_n$. To see that, we note that applying the Liouvillian on a specific $l$-product of $l \leq n$ operators in $\hat{O}_n$ may result in: (i) Addition of a single  spin-half operator to the $l$-product. (ii) There are of order $l$ new operators generated for each $l$-product. 

As a result,  we can only compute numerically a finite set of recurrents, up to order $n_{max}$ and must develop an extrapolation scheme for the higher order recurrents, $ n_{\rm max} < n<\infty $. 

In general, the continued fraction expansion can be formally truncated using a complex termination function $T(z)$
\be 
{C}(z) = 
\cfrac{2}
{z-\cfrac{{\Delta}_1^2}{z-\cfrac{\ddots}{z-\cfrac{{\Delta}_{n_{max}}^2}{z-T(z)}}}}
\label{eq:CFwithT}
\ee 
Clearly, the termination function, $T(z)$, cannot be uniquely inferred given a finite set of low order recurrents. To restrict the functional search space we employ a variational approach. Explicitly, we introduce a complex variational response function, $\tilde{C}(z;\{\alpha_i\})$. The precise choice of the function $\tilde{C}(z;\{\alpha_i\})$ for different observables is discussed in Subsection~\ref{subsec:varfunc},
\be 
\tilde{C}(z;\{\alpha_i\}) = 
\cfrac{2}
{z-\cfrac{\tilde{\Delta}_1^2}{z-\cfrac{\ddots}{z-\cfrac{\tilde{\Delta}_{n_{max}}^2}{z-\tilde{T}(z;\{\alpha_i\})}}}}.
\label{eq:TF}
\ee
In the above equation, we also defined the complex variational termination function $\tilde{T}(z;\{\alpha_i\})$ through the continued fraction.

Our task is to determine the variational parameters ${\alpha_i}$ from the numerically computed set of recurrents $\Delta_n$. Since $\tilde{C}''(\omega;\{\alpha_i\})=-\Im\left[\tilde{C}(z=\omega+i\epsilon;\{\alpha_i\})\right]$ is a known function, and its moments (Eq.~\eqref{eq:moments}) have a closed form, the recurrents $\tilde{\Delta}_n^2$ can be computed numerically to arbitrary precision. This enables us to estimate the variational parameters ${\alpha_i}$ by performing a numerical least square minimization
\be
\chi^2 = \underset{\{\alpha_i\}}{\rm min} \sum\limits_{n=n_{\rm min}}^{n_{\rm max}} \left(\Delta^2_n - \tilde{\Delta}^2_n(\{\alpha\})\right)^2 
\label{eq:LS}
\ee

Empirically, we found that the first few recurrents, $n<n_{\rm min}$, exhibit a transient behavior that deviates from the asymptomatic functional form in Eq.~\eqref{eq:TF}. For this reason, in Eq.~\eqref{eq:LS}, only recurrents with $n\ge n_{\rm min}$ are considered in the numerical fit. In practice, we use $n_{\rm min}=3$ throughout our calculations. We illustrate this procedure in Fig.~\ref{fig:RecExtrapolation}, where we compute the recurrents of the local spin susceptibility for a number of disorder strengths. 

Having determined the variational parameters $\{\alpha_i\}$ we can now invert the continued fraction relation in Eq.~\eqref{eq:TF} to obtain $\tilde{T}(z)$. Finally, in Eq.~\eqref{eq:CFwithT} we substitute ${T}(z)$ in favor of the variational termination function $\tilde{T}(z)$ to form our variational estimate, $C_{\rm VER}(z)$, for the true response function. 

\begin{figure}
	\includegraphics[width=\linewidth]{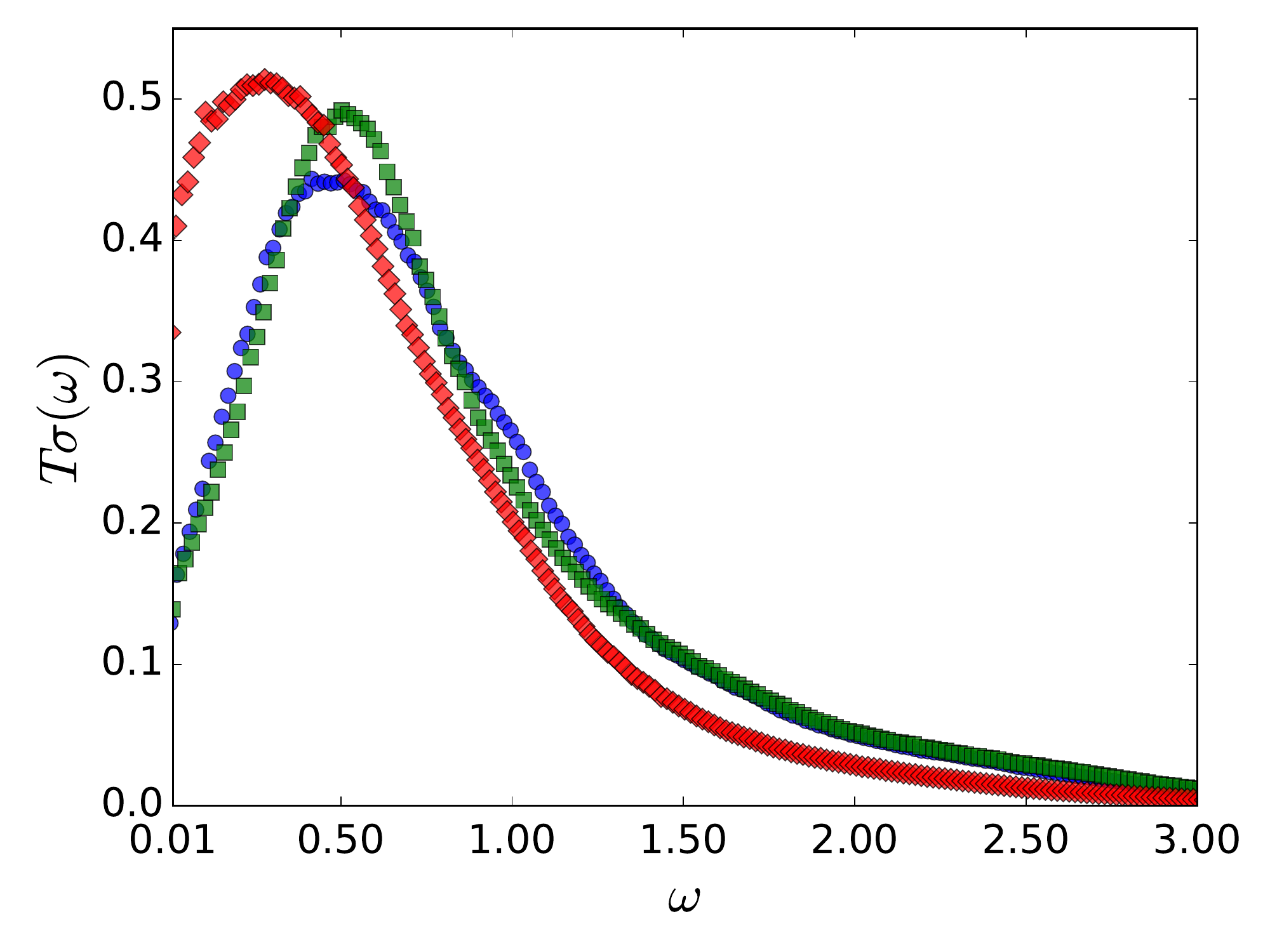}
	\caption{(Color online) ED calculation of $T\sigma(\omega)$ for  $N=14$. Different disorder strengths, (red diamonds) $h/J = 0.5$, (blue circles) $h/J = 0.7$, and (green squares) $h/J = 1.0$ are shown. The structure of the functions resembles a Gaussian with a positive low frequency power law. \label{fig:supEDfunc}}
\end{figure}

The quality of fit and resulting error bars are determined by two criteria: (i) The magnitude of $\chi^2$, the least square fit between the computed and the variational recurrents (Eq.~\eqref{eq:LS}).  
(ii) The convergence of $C_{\rm VER}(\omega) $ with $n_{\rm max}$. This will be discussed in detail in  Subsection~\ref{subsec:errorest}.

\subsection{\label{subsec:varfunc} Choice of Variational Functions} 

For the local spin correlations we modified the variational function suggested by Ref.~\cite{BhomMuller} in the context of clean Heisenberg chains. We use the positive variational functions,
\be
\tilde{C}''=  \left| \omega  \right|^{\beta-1}   \exp{\left\lbrace -\left| \frac{\omega}{\omega_0} \right|^{2/\lambda}\right\rbrace } \left[\left( 1+ \sum_{n=1}^{4}c_n \left| \frac{\omega}{\omega_0} \right|^n \right)^2\right]
\label{eq:Var}
\ee
where $\omega_0, \lambda, \alpha, c_1, c_2,\ldots$ are the  fitted parameters. The rationale for choosing Eq.~\eqref{eq:Var}, is based on physical arguments: 
Known dynamical correlators of similar lattice models (e.g. the $S=1/2$ quantum XY model in one and two dimensions~\cite{BhomMuller,lindnerHCB}) exhibit Gaussian fall-off at high frequencies, with a scale
parametrized by $\omega_0$. We allow for a non-Gaussian fall-off  with a stretch parameter $\lambda$. 
At low frequencies, we allow for an arbitrary power law, which is parametrized by $\beta-1$.  
In the presence of disorder, additional energy scales are expected. Therefore, Eq.~\eqref{eq:Var} can incorporate extra peaks and frequency scales, using higher order polynomial coefficients $c_n, n=1,2,\ldots$.

Insight into the effects of $\omega_0,\beta$, is gained by examining the pure {\em power law $\times$ Gaussian} function (i.e. $c_n=0,~\lambda=1$), whose recurrents are~\cite{viswanath_recursion_book}:
\be
\Delta_{2k}^2 = \omega_0^2  k, \quad \Delta_{2k+1}^2 =   \omega_0^2 ( k+1-\beta/2)
\label{eq:EO}
\ee
Eq.~\eqref{eq:EO} demonstrates  two important points: (i) The average slope at high orders $k\to \infty$ depends on $\omega_0$. (ii) The even-odd alternations at finite $k$, reflect the {\em low frequency} parameter $\beta$. The case of the pure Gaussian ($\beta=1$) has no even-odd alternations, $\Delta_n^2 =  {1\over 2} \omega_0^2  n$.
In Fig.~\ref{fig:RecExtrapolation} we compare the exact recurrents $\Delta_n^2$ of  the local spin autocorrelation function, to those of the VER scheme $\tilde{\Delta}_n^2$, for several different disorder realizations.  Note the even-odd alternation of the recurrents, which signals the low frequency power law singularity of the correlation functions.

\begin{figure} 
    \centering
    \includegraphics[width=1\columnwidth]{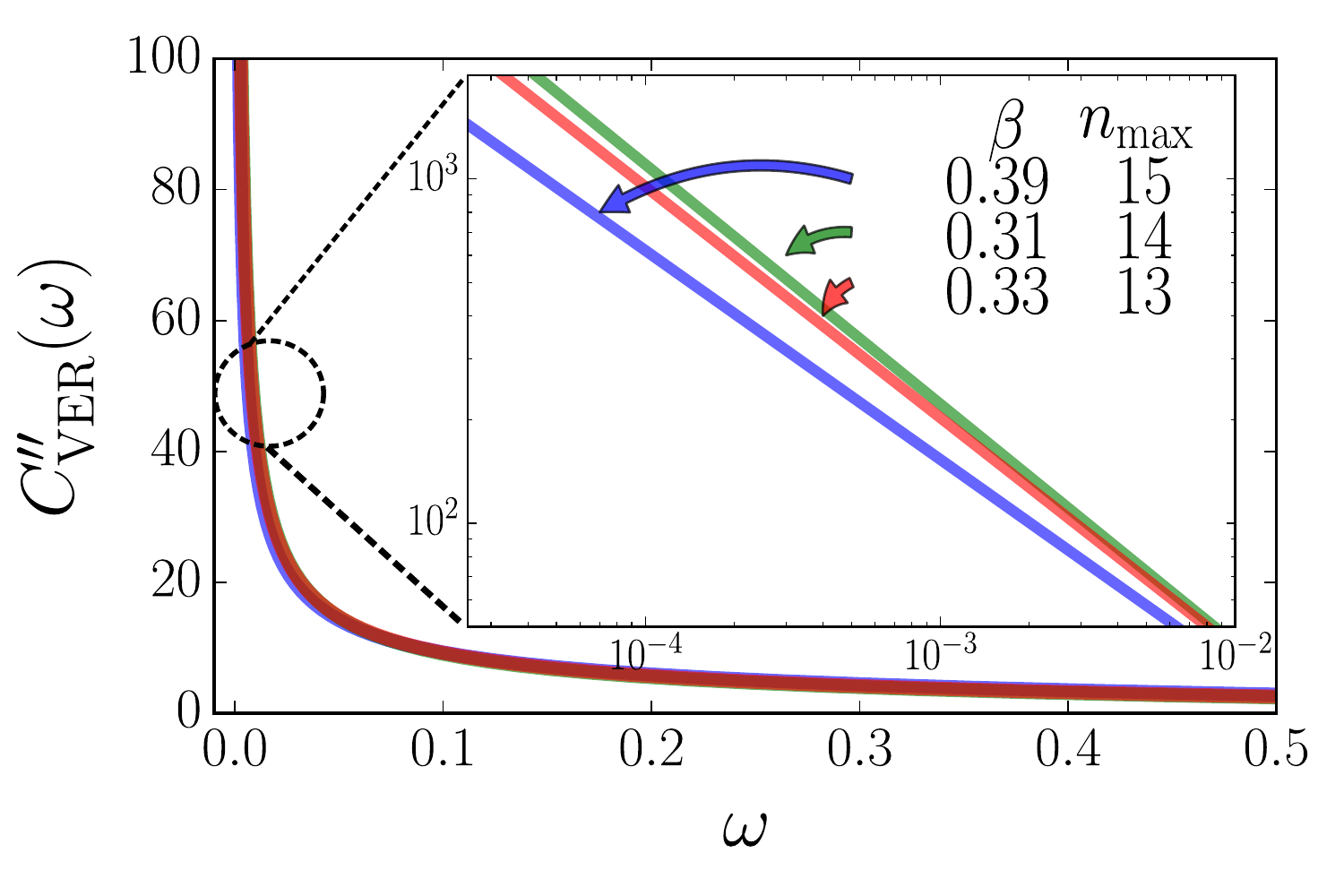}
    \caption{(Color online) Local spin correlation function with disorder strength $h=0.5$. VER scheme (See text) results with increasing number of recurrents $n_{\rm max}$. Inset: Extraction of exponent $\beta$.
 \label{fig:VERExtrapolation}}      
\end{figure}

To obtain an educated guess for the functional form of the variational ansatz (Eq.~\eqref{eq:TF}) for the AC conductivity we use ED on small systems, up to $N=14$ sites. This ED calculation involves the full Hilbert space in order to comply with VER calculations which are not restricted to the $\sum_i S_i^z=0$ sector. In Fig.~\ref{fig:supEDfunc}, we depict $T\sigma(\omega)$ for different disorder realizations. We see that all curves can be modeled using a power law multiplying a Gaussian. In addition, in certain disorder realizations, we notice that the dynamical conductivity displays an additional shoulder at finite frequency. This effect is taken into account by adding a symmetric Gaussian whose center is shifted (Eq.~\eqref{eq:TS1}). The considerations above led us to this variational function:
\bea
T \tilde{\sigma} \left(\omega\right) &=& C \left| \frac{\omega}{\omega_0} \right|^\alpha \left( \exp{ \left( -\left| \frac{\omega}{\omega_0} \right| ^{2/\lambda} \right)}  \right. \nonumber \\
&~&+ \left. B \exp{\left(- \left(\frac{\omega+\omega_1}{\omega_2} \right)^{2} \right)} + \omega_1\rightarrow-\omega_1 \right),\nonumber\\
\label{eq:sigma-var}
\eea
In the clean limit, a spurious zero frequency delta function appears in the continued fraction  because of the inclusion of the $n=m$ terms in Eq.~\eqref{eq:TS}.
(A spurious zero frequency delta-function appears in ED calculations for a different reason~\cite{AffleckHighTDiff}).
At finite but weak disorder, the spurious delta function becomes negligible, which enables a good fit to our variational function Eq.~\eqref{eq:sigma-var}.

\begin{figure} 
    \centering
    \includegraphics[width=1\columnwidth]{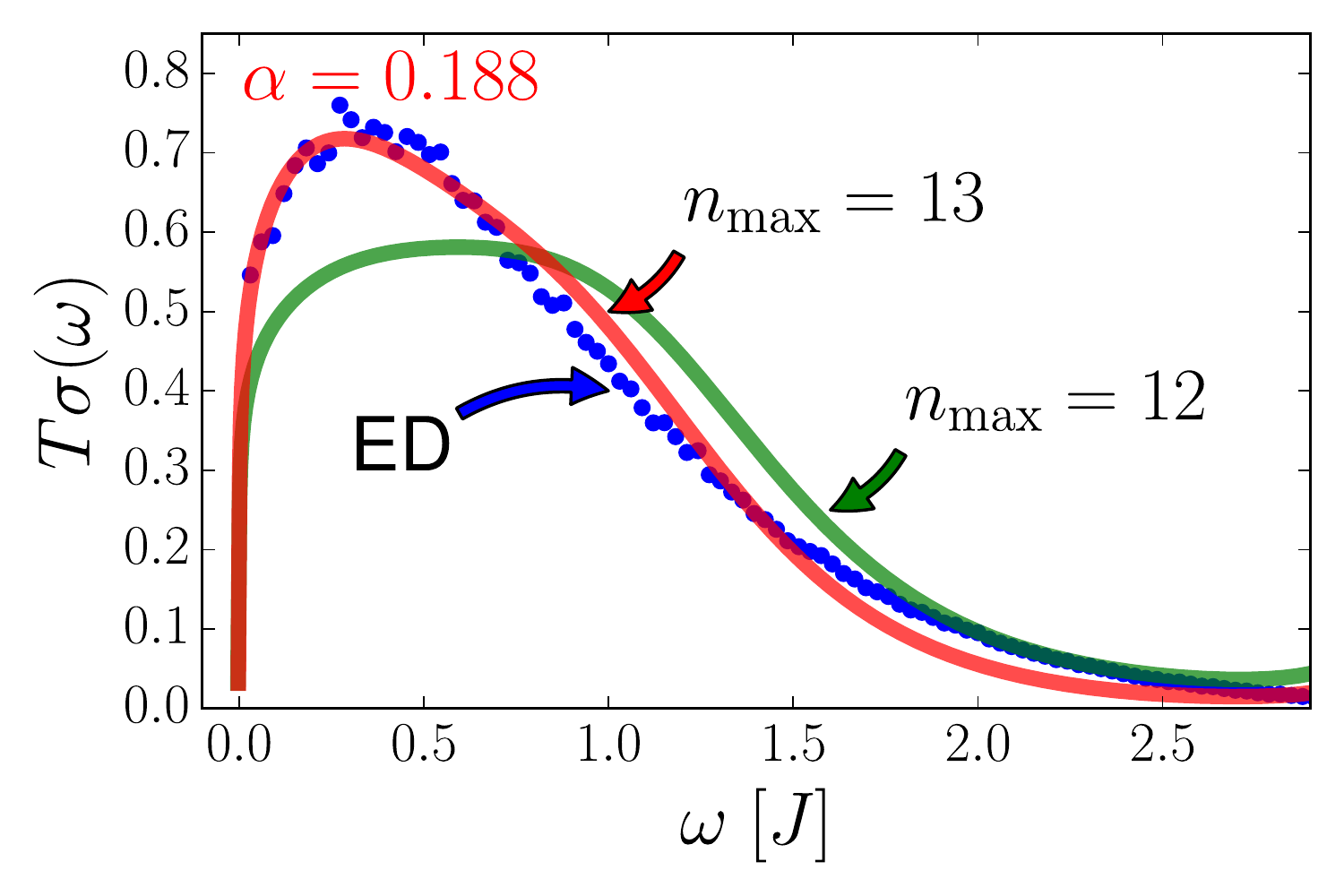}
    \caption{(Color online) AC conductivity of disordered Heisenberg chain of $N=14$ sites. The plot represents a single realization of disorder with $h=0.5$. Exact diagonalization of Eq.~\eqref{eq:TS} (blue circles) is compared to the VER calculations (solid lines) for different values of $n_{\rm max}$.
    Good agreement is reached for $n_{\rm max} =13$.  $\alpha$  represents the low frequency power law. \label{fig:EDcompare}}     
\end{figure}

\section{\label{sec:results} Results}
\subsection{\label{subsec:verres} Variational Extrapolation of Recurrents}

We have computed the operator traces in Eq.~\eqref{eq:mu2k} up to order $k=n_{\rm max}$ for the local spin correlations, and the conductivity. 
For the clean Heisenberg model, with $n_{\rm max}=19$, the  local spin correlations exhibit a low frequency power law, 
\be
C_{\rm Heis}''(\omega) \sim \omega^{-0.459 \pm 0.065}
\ee

For finite disorder, the values of the exponent $\beta$ extracted from $C''(\omega) \sim \omega^{\beta-1}$, are presented in Fig.~\ref{fig:PhaseDiagram}. For each disorder strength, we use $10^3$ realizations and $n_{\rm max} = 15$. 
The AC conductivity  is defined on a lattice of size $N=30$, while recurrents are computed up to order $n_{\rm max}=12$. The distribution of these exponents is discussed in Subsection~\ref{subsec:errorest}.

We plot the disorder-averaged  $\alpha,\beta$ in Fig.~\ref{fig:PhaseDiagram}. 
Contrary to previous observations~\cite{AgarwalSubDiff}, but with some agreement with Ref.~\cite{Luitz2015} and larger scale ED results up to $N=22$ (see Sec.~\ref{subsec:edres} for full range of disorder strengths), we find no diffusion $(\beta={1\over 2}, \alpha=0)$ for disorders with $0< h \le  1.5$.
Instead, we find a single sub-diffusive phase ($\beta<0.5,~\alpha>0$) which begins at arbitrarily weak disorder. 
We are unable to directly probe the MBL transition using VER, since beyond $h \gtrsim 1.5$ the relative errors in extracting $\alpha, \beta$ are greater than $25\%$. 
The critical disorder was estimated to be  $h_{\rm MBL} \approx 3.7$~\cite{AgarwalSubDiff,AletL22,Luitz2015} and is consistent with a  naive extrapolation of our results  to stronger disorder.
  
The sum  $\alpha+2\beta$ is also plotted in Fig.~1. The scaling hypothesis  (Eq.~\eqref{eq:scaling}) is verified in the stronger disorder regime.  
However, at weak disorder we note a systematic deviation 
from unity. We have no explanation for the deviation from scaling in this regime. We leave open the possibility that it might be an artifact of
systematic errors in the VER scheme, and  insufficient averaging over disorder realizations.

\begin{figure}[t!]
	\includegraphics[width=\linewidth]{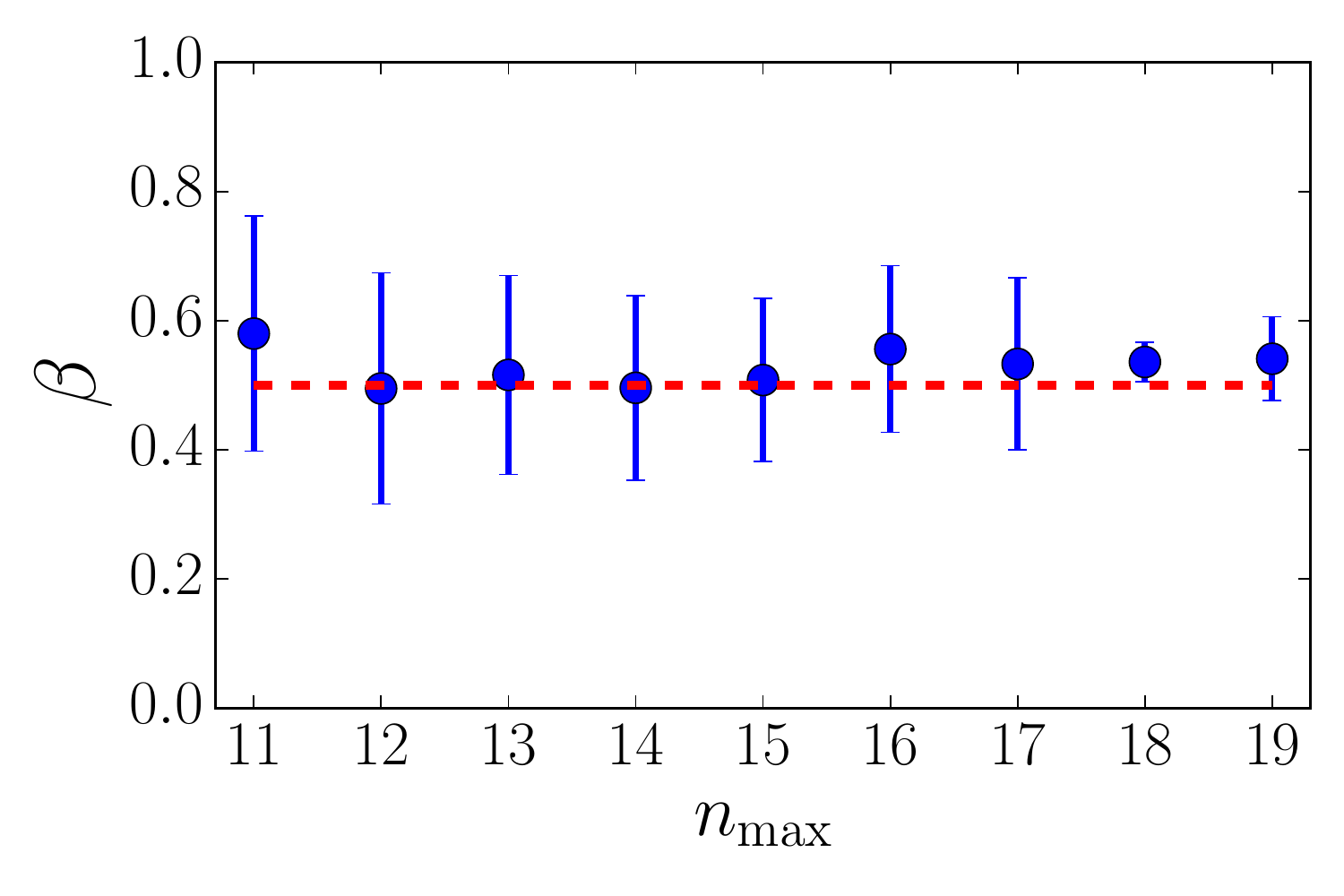}
	\caption{(Color online) $\beta$ (blue circles) of the Clean Heisenberg Hamiltonian as extracted from varying the number of recurrents used in the VER scheme, $n_{\rm max}$. The error bars are the uncertainty intervals of the least square procedure. \label{fig:CleanRec}}
\end{figure}

\subsection{\label{subsec:errorest} Convergence and Error Estimation}

We investigate the stability of the VER scheme by increasing  the number of recurrents $n_{\rm max}$. 
Convergence of the local spin correlation function and the AC conductivity are shown in  Fig.~\ref{fig:VERExtrapolation} and Fig.~\ref{fig:EDcompare}, respectively. 
For the latter, the VER extrapolation is compared directly with ED up to $N=14$ and we observe that 13 recurrents are sufficient to recover the exact result.  The finite DC conductivity seen in the ED calculation (Fig.\ref{fig:supEDfunc}, \ref{fig:EDcompare}) is an artifact arising from the delta function broadening parameter in Eq.~\eqref{eq:TS}. 

The dynamical response of the clean Heisenberg limit is explored in Fig.~\ref{fig:CleanRec}. We deduce the diffusive exponent value $\beta = 0.5$ from the VER analysis of the first $19$ recurrents.
The error in the exponents, $\alpha$ and $\beta$, is estimated by the least square fit $\chi^2$ (Eq.~\eqref{eq:LS}), and is found to be larger than the statistical error arising from disorder averaging.

\begin{figure}[t]
	\includegraphics[width=\linewidth]{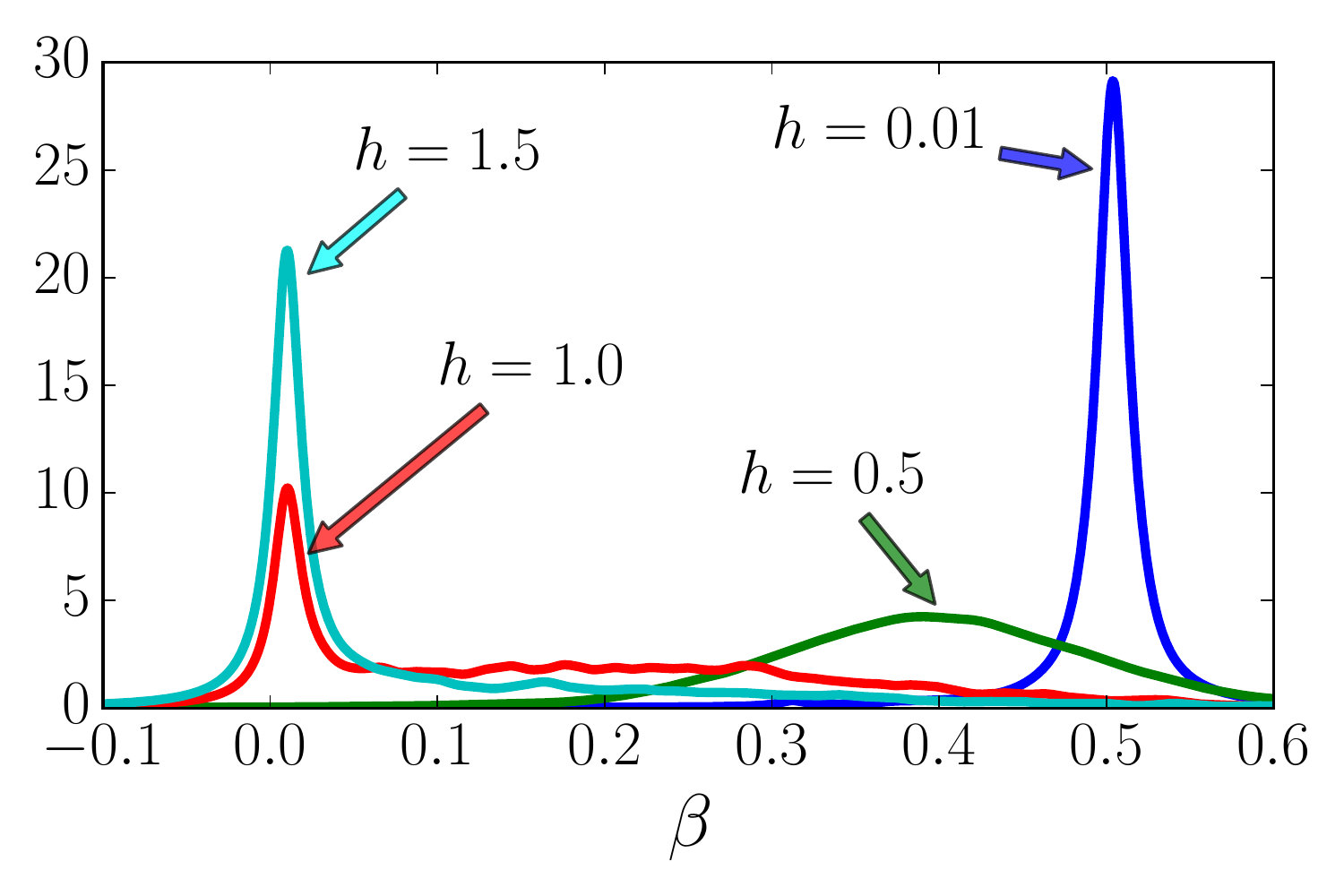}
	\caption{(Color online) Probability distribution of $\beta$ for different disorder strengths. We witness a flow of weight towards $\beta \rightarrow 0$ as disorder is increased. An error estimate can be extracted from the width of the distribution.  \label{fig:BetaHist}}
\end{figure}

It is instructive to explore the probability density function (PDF) of the extracted exponents and  Fig.~\ref{fig:BetaHist}  depicts the numerically computed PDF of   $\beta$ for different values of disorder. We note that in all cases the probability is peaked about its center. The width of the histogram is expected to decrease with system size and eventually vanish in the thermodynamic limit.
We also note that, as expected, with increasing disorder, the weight flows toward the localized regime, i.e. $\beta \rightarrow 0$.

\subsection{\label{subsec:edres} Exact Diagonalization}

To benchmark the prediction of the VER scheme, we solve for Eq.\eqref{eq:Ham} using exact diagonalizaion (ED) on  large chains of size up to $N=22$. We perform full diagonalization of the $\sum_i S^z_i=0$ sector using a layered shift and invert spectral transformation. We utilize the SLEPc (Scalable Library for Eigenvalue Problem Computations) library to apply the transformation in a parallelized way and compute the time dependent local spin correlation function. The power law exponent, $\beta$, is fitted using a time window that begins after the initial transient, and ends before the appearance of finite size effects at long times, seen as flattening of the response function (see Fig.~\ref{fig:L22ED}). We note that the ED results predicts sub-diffusive transport even at small disorder strengths, $h/J \sim 0.5$. This is in qualitative agreement with the VER approach. 

We emphasize that it is essential to study relatively large system sizes in order to probe the nature of the sub-diffusive Griffiths phase. For  smaller system sizes, $N \le 14$, one finds an exponent $\beta > 0.5$ for small disorders $h < 1.0$. This is a finite size artifact that arises from fitting the transient in the relatively short time windows before finite size features appear. Observing the scaling of $C(t)$ as a function of system size reveals a systematic shift to lower exponents consistent with the VER approach and sub-diffusion at even small disorders.

\section{\label{sec:discusion} Discussion}

The continued fraction describes, in principle, the infinite size lattice. It  describes equally well the high and  low frequency regimes.
We have seen, in Eq.~\eqref{eq:EO}, that
the asymptotic low frequency power law determines the alternation between the even and odd order recurrents. This can be observed even 
by looking at the behavior of the low  order recurrents in Fig.~\ref{fig:RecExtrapolation}.

In this work, similarly to previous studies \cite{AgarwalSubDiff,BarLevSubDiff,Luitz2015,OganesyanHighT}, we focus solely on the infinite temperature limit. Due to the finite on-site Hilbert space, the energy density is bounded. As a result, at finite but large temperature, $T\gg J$, the Kubo formula would receive relative corrections of order  $ \mathcal{O}(J/T)^2$.  Importantly, universal properties and in particular the sub-diffusive nature of transport, are expected to remain valid. Finite temperature corrections to transport can, in principle, be incorporated into the VER calculation. We leave this interesting line of research to future studies.

The VER method however has its limitations. Its accuracy depends on the  choice of the variational family of trial functions.
 This  can be improved by adding more parameters.  
 Confidence in the resulting correlation function is increased by testing for convergence with $n_{\max}$.  
 However, for interacting models, the cost of computing high order recurrents increases exponentially (or faster) with  $n_{\rm max}$. 
Hence it is reassuring to find  examples,  such as the clean Heisenberg limit,  where  $ C_{\rm VER}$ converges rapidly with $n_{\rm max}$, as shown in Fig.~\ref{fig:VERExtrapolation}.

 {\em How does VER  compare to existing numerical approaches?} The computational costs of ED also increase exponentially, but with the lattice size. Hence, ED is limited by
boundary effects, which dominate the long time behavior of the response functions.
This is especially problematic in the delocalized, weak disorder regime. For example, ED calculations run into long time saturation especially in the weak disorder thermal phase (see inset of Fig.~\ref{fig:L22ED}). Therefore ED and VER are complementary and suitable in different regimes - ED in the MBL phase, and VER for the ergodic phase. 

The Numerical Linked Cluster (NLC) method~\cite{RajivNLC} addresses the thermodynamic limit by extrapolation. It is  interesting to note  that NLC works better in the localized regime while VER works in the delocalized phase, such that a combination of the two can be particularly powerful. 

Our primary conclusion is the {\em absence of diffusion for the disordered Heisenberg model} at any finite disorder strength.
This is indicated by  $\beta<0.5$ of the spin relaxation, and by $\alpha>0$ of the conductivity.
It is interesting to ask if this result depends on (1) Dimensionality - is there a diffusive phase in two and higher dimensions? and  (ii) Integrability of the clean limit - Could the strong sensitivity to weak disorder be related to the extensive number of local conserved operators in the clean Heisenberg model?
Answers to these questions demand additional studies.
Based on the results above, we find  the continued fraction VER method to be a promising route to address these questions.

\section*{Acknowledgments}

We thank Dan Arovas, Netanel Lindner, Daniel Podolsky, Joel E. Moore and  Ehud Altman for beneficial discussions. We are grateful to David Cohen, the administrator of the ATLAS-Technion grid project for his extensive support. SG received support from the Simons Investigators Program and the California Institute of Quantum Emulation. NYY acknowledges the Miller Institute for Basic Research in Science.  AA gratefully acknowledges support from the US-Israel Binational Science Foundation grant 2012233 and the Israel Science Foundation and thank the Aspen Center for Physics, supported by the NSF-PHY-1066293, for its hospitality. 

\appendix
	\section{Recurrents to Moments and vice versa}
	The moments of the variational response function (Eq.~\eqref{eq:TF}) are known in closed from. The recurrents can then be computed following the method suggested in Ref.~\cite{viswanath_recursion_book} for converting moments to recurrents.
	
	{\em Moments to recurrents - } given a set of moments $\mu_{2k},~k=0,\ldots,n_{\rm max}$ with $\mu_0=1$. The recurrents $\Delta^2_n \equiv \mu_{2n}^{(n)}$ are extracted from
	\be
	\mu_{2k}^{(n)} = \frac{\mu_{2k}^{(n-1)}}{\Delta^2_{n-1}} - \frac{\mu_{2(k-1)}^{(n-2)}}{\Delta^2_{n-2}}
	\label{eq:momentstorec}
	\ee
	for $n=1,\ldots,n_{\rm max}$ and $k = m,\ldots,n_{\rm max}$. With initial values $\mu_{2k}^{(0)}=\mu_{2k}$, $\Delta^2_{-1}=\Delta^2_0=1$, and $\mu_{2k}^{(-1)}=0$.
	
	For completeness we add the inverse transformation,
	{\em Recurrents to moments - } given a set of recurrents $\Delta^2_n,~n=1,\ldots,n_{\rm max}$ and $\Delta^2_{-1}=\Delta^2_0=1$, the moments $\mu_{2n} \equiv \mu_{2n}^{(0)}$ are extracted from
	\be
	\mu_{2k}^{(n-1)} = \mu_{2k}^{(n)}\Delta^2_{n-1} + \mu_{2(k-1)}^{(n-2)}\frac{\Delta^2_{n-1}}{\Delta^2_{n-2}}
	\ee
	with $\mu_{2n}^{(n)} = \Delta^2_n$. For $n=k,k-1,\ldots,1$, $k=1,\ldots,n_{\rm max}$, and with initial values $\mu_{2k}^{(-1)}=0$.


\begin{thebibliography}{38}%
\makeatletter
\providecommand \@ifxundefined [1]{%
 \@ifx{#1\undefined}
}%
\providecommand \@ifnum [1]{%
 \ifnum #1\expandafter \@firstoftwo
 \else \expandafter \@secondoftwo
 \fi
}%
\providecommand \@ifx [1]{%
 \ifx #1\expandafter \@firstoftwo
 \else \expandafter \@secondoftwo
 \fi
}%
\providecommand \natexlab [1]{#1}%
\providecommand \enquote  [1]{``#1''}%
\providecommand \bibnamefont  [1]{#1}%
\providecommand \bibfnamefont [1]{#1}%
\providecommand \citenamefont [1]{#1}%
\providecommand \href@noop [0]{\@secondoftwo}%
\providecommand \href [0]{\begingroup \@sanitize@url \@href}%
\providecommand \@href[1]{\@@startlink{#1}\@@href}%
\providecommand \@@href[1]{\endgroup#1\@@endlink}%
\providecommand \@sanitize@url [0]{\catcode `\\12\catcode `\$12\catcode
  `\&12\catcode `\#12\catcode `\^12\catcode `\_12\catcode `\%12\relax}%
\providecommand \@@startlink[1]{}%
\providecommand \@@endlink[0]{}%
\providecommand \url  [0]{\begingroup\@sanitize@url \@url }%
\providecommand \@url [1]{\endgroup\@href {#1}{\urlprefix }}%
\providecommand \urlprefix  [0]{URL }%
\providecommand \Eprint [0]{\href }%
\providecommand \doibase [0]{http://dx.doi.org/}%
\providecommand \selectlanguage [0]{\@gobble}%
\providecommand \bibinfo  [0]{\@secondoftwo}%
\providecommand \bibfield  [0]{\@secondoftwo}%
\providecommand \translation [1]{[#1]}%
\providecommand \BibitemOpen [0]{}%
\providecommand \bibitemStop [0]{}%
\providecommand \bibitemNoStop [0]{.\EOS\space}%
\providecommand \EOS [0]{\spacefactor3000\relax}%
\providecommand \BibitemShut  [1]{\csname bibitem#1\endcsname}%
\let\auto@bib@innerbib\@empty
\bibitem [{\citenamefont {Anderson}(1958)}]{anderson_loc}%
  \BibitemOpen
  \bibfield  {author} {\bibinfo {author} {\bibfnamefont {P.~W.}\ \bibnamefont
  {Anderson}},\ }\href {\doibase 10.1103/PhysRev.109.1492} {\bibfield
  {journal} {\bibinfo  {journal} {Phys. Rev.}\ }\textbf {\bibinfo {volume}
  {109}},\ \bibinfo {pages} {1492} (\bibinfo {year} {1958})}\BibitemShut
  {NoStop}%
\bibitem [{\citenamefont {{Nandkishore}}\ and\ \citenamefont
  {{Huse}}(2015)}]{MBLReview}%
  \BibitemOpen
  \bibfield  {author} {\bibinfo {author} {\bibfnamefont {R.}~\bibnamefont
  {{Nandkishore}}}\ and\ \bibinfo {author} {\bibfnamefont {D.~A.}\ \bibnamefont
  {{Huse}}},\ }\href {\doibase 10.1146/annurev-conmatphys-031214-014726}
  {\bibfield  {journal} {\bibinfo  {journal} {Annual Review of Condensed Matter
  Physics}\ }\textbf {\bibinfo {volume} {6}},\ \bibinfo {pages} {15} (\bibinfo
  {year} {2015})}\BibitemShut {NoStop}%
\bibitem [{\citenamefont {Basko}\ \emph {et~al.}(2006)\citenamefont {Basko},
  \citenamefont {Aleiner},\ and\ \citenamefont {Altshuler}}]{BaskoMBL}%
  \BibitemOpen
  \bibfield  {author} {\bibinfo {author} {\bibfnamefont {D.}~\bibnamefont
  {Basko}}, \bibinfo {author} {\bibfnamefont {I.}~\bibnamefont {Aleiner}}, \
  and\ \bibinfo {author} {\bibfnamefont {B.}~\bibnamefont {Altshuler}},\ }\href
  {\doibase http://dx.doi.org/10.1016/j.aop.2005.11.014} {\bibfield  {journal}
  {\bibinfo  {journal} {Annals of Physics}\ }\textbf {\bibinfo {volume}
  {321}},\ \bibinfo {pages} {1126 } (\bibinfo {year} {2006})}\BibitemShut
  {NoStop}%
\bibitem [{\citenamefont {Oganesyan}\ and\ \citenamefont
  {Huse}(2007)}]{OganesyanHighT}%
  \BibitemOpen
  \bibfield  {author} {\bibinfo {author} {\bibfnamefont {V.}~\bibnamefont
  {Oganesyan}}\ and\ \bibinfo {author} {\bibfnamefont {D.~A.}\ \bibnamefont
  {Huse}},\ }\href {\doibase 10.1103/PhysRevB.75.155111} {\bibfield  {journal}
  {\bibinfo  {journal} {Phys. Rev. B}\ }\textbf {\bibinfo {volume} {75}},\
  \bibinfo {pages} {155111} (\bibinfo {year} {2007})}\BibitemShut {NoStop}%
\bibitem [{\citenamefont {\v{Z}nidari\v{c}}\ \emph {et~al.}(2008)\citenamefont
  {\v{Z}nidari\v{c}}, \citenamefont {Prosen},\ and\ \citenamefont
  {Prelov\v{s}ek}}]{ZnidariED}%
  \BibitemOpen
  \bibfield  {author} {\bibinfo {author} {\bibfnamefont {M.}~\bibnamefont
  {\v{Z}nidari\v{c}}}, \bibinfo {author} {\bibfnamefont {T.}~\bibnamefont
  {Prosen}}, \ and\ \bibinfo {author} {\bibfnamefont {P.}~\bibnamefont
  {Prelov\v{s}ek}},\ }\href {\doibase 10.1103/PhysRevB.77.064426} {\bibfield
  {journal} {\bibinfo  {journal} {Phys. Rev. B}\ }\textbf {\bibinfo {volume}
  {77}},\ \bibinfo {pages} {064426} (\bibinfo {year} {2008})}\BibitemShut
  {NoStop}%
\bibitem [{\citenamefont {{Luitz}}\ \emph {et~al.}(2015)\citenamefont
  {{Luitz}}, \citenamefont {{Laflorencie}},\ and\ \citenamefont
  {{Alet}}}]{Luitz2015}%
  \BibitemOpen
  \bibfield  {author} {\bibinfo {author} {\bibfnamefont {D.~J.}\ \bibnamefont
  {{Luitz}}}, \bibinfo {author} {\bibfnamefont {N.}~\bibnamefont
  {{Laflorencie}}}, \ and\ \bibinfo {author} {\bibfnamefont {F.}~\bibnamefont
  {{Alet}}},\ }\href@noop {} {\bibfield  {journal} {\bibinfo  {journal} {ArXiv
  e-prints}\ } (\bibinfo {year} {2015})},\ \Eprint
  {http://arxiv.org/abs/1511.05141} {arXiv:1511.05141 [cond-mat.dis-nn]}
  \BibitemShut {NoStop}%
\bibitem [{\citenamefont {Devakul}\ and\ \citenamefont
  {Singh}(2015)}]{RajivNLC}%
  \BibitemOpen
  \bibfield  {author} {\bibinfo {author} {\bibfnamefont {T.}~\bibnamefont
  {Devakul}}\ and\ \bibinfo {author} {\bibfnamefont {R.~R.~P.}\ \bibnamefont
  {Singh}},\ }\href {\doibase 10.1103/PhysRevLett.115.187201} {\bibfield
  {journal} {\bibinfo  {journal} {Phys. Rev. Lett.}\ }\textbf {\bibinfo
  {volume} {115}},\ \bibinfo {pages} {187201} (\bibinfo {year}
  {2015})}\BibitemShut {NoStop}%
\bibitem [{\citenamefont {Bardarson}\ \emph {et~al.}(2012)\citenamefont
  {Bardarson}, \citenamefont {Pollmann},\ and\ \citenamefont
  {Moore}}]{BardarsonLogEnt}%
  \BibitemOpen
  \bibfield  {author} {\bibinfo {author} {\bibfnamefont {J.~H.}\ \bibnamefont
  {Bardarson}}, \bibinfo {author} {\bibfnamefont {F.}~\bibnamefont {Pollmann}},
  \ and\ \bibinfo {author} {\bibfnamefont {J.~E.}\ \bibnamefont {Moore}},\
  }\href {\doibase 10.1103/PhysRevLett.109.017202} {\bibfield  {journal}
  {\bibinfo  {journal} {Phys. Rev. Lett.}\ }\textbf {\bibinfo {volume} {109}},\
  \bibinfo {pages} {017202} (\bibinfo {year} {2012})}\BibitemShut {NoStop}%
\bibitem [{\citenamefont {Serbyn}\ \emph
  {et~al.}(2013{\natexlab{a}})\citenamefont {Serbyn}, \citenamefont
  {Papi\ifmmode~\acute{c}\else \'{c}\fi{}},\ and\ \citenamefont
  {Abanin}}]{SerbynSlow}%
  \BibitemOpen
  \bibfield  {author} {\bibinfo {author} {\bibfnamefont {M.}~\bibnamefont
  {Serbyn}}, \bibinfo {author} {\bibfnamefont {Z.}~\bibnamefont
  {Papi\ifmmode~\acute{c}\else \'{c}\fi{}}}, \ and\ \bibinfo {author}
  {\bibfnamefont {D.~A.}\ \bibnamefont {Abanin}},\ }\href {\doibase
  10.1103/PhysRevLett.110.260601} {\bibfield  {journal} {\bibinfo  {journal}
  {Phys. Rev. Lett.}\ }\textbf {\bibinfo {volume} {110}},\ \bibinfo {pages}
  {260601} (\bibinfo {year} {2013}{\natexlab{a}})}\BibitemShut {NoStop}%
\bibitem [{\citenamefont {Serbyn}\ \emph
  {et~al.}(2013{\natexlab{b}})\citenamefont {Serbyn}, \citenamefont
  {Papi\ifmmode~\acute{c}\else \'{c}\fi{}},\ and\ \citenamefont
  {Abanin}}]{SerbynLocalIntegrals}%
  \BibitemOpen
  \bibfield  {author} {\bibinfo {author} {\bibfnamefont {M.}~\bibnamefont
  {Serbyn}}, \bibinfo {author} {\bibfnamefont {Z.}~\bibnamefont
  {Papi\ifmmode~\acute{c}\else \'{c}\fi{}}}, \ and\ \bibinfo {author}
  {\bibfnamefont {D.~A.}\ \bibnamefont {Abanin}},\ }\href {\doibase
  10.1103/PhysRevLett.111.127201} {\bibfield  {journal} {\bibinfo  {journal}
  {Phys. Rev. Lett.}\ }\textbf {\bibinfo {volume} {111}},\ \bibinfo {pages}
  {127201} (\bibinfo {year} {2013}{\natexlab{b}})}\BibitemShut {NoStop}%
\bibitem [{\citenamefont {Huse}\ \emph {et~al.}(2014)\citenamefont {Huse},
  \citenamefont {Nandkishore},\ and\ \citenamefont {Oganesyan}}]{HusePheno}%
  \BibitemOpen
  \bibfield  {author} {\bibinfo {author} {\bibfnamefont {D.~A.}\ \bibnamefont
  {Huse}}, \bibinfo {author} {\bibfnamefont {R.}~\bibnamefont {Nandkishore}}, \
  and\ \bibinfo {author} {\bibfnamefont {V.}~\bibnamefont {Oganesyan}},\ }\href
  {\doibase 10.1103/PhysRevB.90.174202} {\bibfield  {journal} {\bibinfo
  {journal} {Phys. Rev. B}\ }\textbf {\bibinfo {volume} {90}},\ \bibinfo
  {pages} {174202} (\bibinfo {year} {2014})}\BibitemShut {NoStop}%
\bibitem [{\citenamefont {Vosk}\ and\ \citenamefont {Altman}(2013)}]{VoskIM}%
  \BibitemOpen
  \bibfield  {author} {\bibinfo {author} {\bibfnamefont {R.}~\bibnamefont
  {Vosk}}\ and\ \bibinfo {author} {\bibfnamefont {E.}~\bibnamefont {Altman}},\
  }\href {\doibase 10.1103/PhysRevLett.110.067204} {\bibfield  {journal}
  {\bibinfo  {journal} {Phys. Rev. Lett.}\ }\textbf {\bibinfo {volume} {110}},\
  \bibinfo {pages} {067204} (\bibinfo {year} {2013})}\BibitemShut {NoStop}%
\bibitem [{\citenamefont {{Imbrie}}(2014)}]{imbrie2014many}%
  \BibitemOpen
  \bibfield  {author} {\bibinfo {author} {\bibfnamefont {J.~Z.}\ \bibnamefont
  {{Imbrie}}},\ }\href@noop {} {\bibfield  {journal} {\bibinfo  {journal}
  {ArXiv e-prints}\ } (\bibinfo {year} {2014})},\ \Eprint
  {http://arxiv.org/abs/1403.7837} {arXiv:1403.7837 [math-ph]} \BibitemShut
  {NoStop}%
\bibitem [{\citenamefont {Schreiber}\ \emph {et~al.}(2015)\citenamefont
  {Schreiber}, \citenamefont {Hodgman}, \citenamefont {Bordia}, \citenamefont
  {Lüschen}, \citenamefont {Fischer}, \citenamefont {Vosk}, \citenamefont
  {Altman}, \citenamefont {Schneider},\ and\ \citenamefont
  {Bloch}}]{SchreiberMBLColdAtoms}%
  \BibitemOpen
  \bibfield  {author} {\bibinfo {author} {\bibfnamefont {M.}~\bibnamefont
  {Schreiber}}, \bibinfo {author} {\bibfnamefont {S.~S.}\ \bibnamefont
  {Hodgman}}, \bibinfo {author} {\bibfnamefont {P.}~\bibnamefont {Bordia}},
  \bibinfo {author} {\bibfnamefont {H.~P.}\ \bibnamefont {Lüschen}}, \bibinfo
  {author} {\bibfnamefont {M.~H.}\ \bibnamefont {Fischer}}, \bibinfo {author}
  {\bibfnamefont {R.}~\bibnamefont {Vosk}}, \bibinfo {author} {\bibfnamefont
  {E.}~\bibnamefont {Altman}}, \bibinfo {author} {\bibfnamefont
  {U.}~\bibnamefont {Schneider}}, \ and\ \bibinfo {author} {\bibfnamefont
  {I.}~\bibnamefont {Bloch}},\ }\href {\doibase 10.1126/science.aaa7432}
  {\bibfield  {journal} {\bibinfo  {journal} {Science}\ }\textbf {\bibinfo
  {volume} {349}},\ \bibinfo {pages} {842} (\bibinfo {year}
  {2015})}\BibitemShut {NoStop}%
\bibitem [{\citenamefont {Kondov}\ \emph {et~al.}(2015)\citenamefont {Kondov},
  \citenamefont {McGehee}, \citenamefont {Xu},\ and\ \citenamefont
  {DeMarco}}]{DeMarco}%
  \BibitemOpen
  \bibfield  {author} {\bibinfo {author} {\bibfnamefont {S.~S.}\ \bibnamefont
  {Kondov}}, \bibinfo {author} {\bibfnamefont {W.~R.}\ \bibnamefont {McGehee}},
  \bibinfo {author} {\bibfnamefont {W.}~\bibnamefont {Xu}}, \ and\ \bibinfo
  {author} {\bibfnamefont {B.}~\bibnamefont {DeMarco}},\ }\href {\doibase
  10.1103/PhysRevLett.114.083002} {\bibfield  {journal} {\bibinfo  {journal}
  {Phys. Rev. Lett.}\ }\textbf {\bibinfo {volume} {114}},\ \bibinfo {pages}
  {083002} (\bibinfo {year} {2015})}\BibitemShut {NoStop}%
\bibitem [{\citenamefont {{Smith}}\ \emph {et~al.}(2015)\citenamefont
  {{Smith}}, \citenamefont {{Lee}}, \citenamefont {{Richerme}}, \citenamefont
  {{Neyenhuis}}, \citenamefont {{Hess}}, \citenamefont {{Hauke}}, \citenamefont
  {{Heyl}}, \citenamefont {{Huse}},\ and\ \citenamefont
  {{Monroe}}}]{MBLTrappedIons}%
  \BibitemOpen
  \bibfield  {author} {\bibinfo {author} {\bibfnamefont {J.}~\bibnamefont
  {{Smith}}}, \bibinfo {author} {\bibfnamefont {A.}~\bibnamefont {{Lee}}},
  \bibinfo {author} {\bibfnamefont {P.}~\bibnamefont {{Richerme}}}, \bibinfo
  {author} {\bibfnamefont {B.}~\bibnamefont {{Neyenhuis}}}, \bibinfo {author}
  {\bibfnamefont {P.~W.}\ \bibnamefont {{Hess}}}, \bibinfo {author}
  {\bibfnamefont {P.}~\bibnamefont {{Hauke}}}, \bibinfo {author} {\bibfnamefont
  {M.}~\bibnamefont {{Heyl}}}, \bibinfo {author} {\bibfnamefont {D.~A.}\
  \bibnamefont {{Huse}}}, \ and\ \bibinfo {author} {\bibfnamefont
  {C.}~\bibnamefont {{Monroe}}},\ }\href {http://arxiv.org/abs/1508.07026}
  {\bibfield  {journal} {\bibinfo  {journal} {ArXiv e-prints}\ } (\bibinfo
  {year} {2015})},\ \Eprint {http://arxiv.org/abs/1508.07026} {arXiv:1508.07026
  [quant-ph]} \BibitemShut {NoStop}%
\bibitem [{\citenamefont {Vosk}\ \emph {et~al.}(2015)\citenamefont {Vosk},
  \citenamefont {Huse},\ and\ \citenamefont {Altman}}]{VoskRSRG}%
  \BibitemOpen
  \bibfield  {author} {\bibinfo {author} {\bibfnamefont {R.}~\bibnamefont
  {Vosk}}, \bibinfo {author} {\bibfnamefont {D.~A.}\ \bibnamefont {Huse}}, \
  and\ \bibinfo {author} {\bibfnamefont {E.}~\bibnamefont {Altman}},\ }\href
  {\doibase 10.1103/PhysRevX.5.031032} {\bibfield  {journal} {\bibinfo
  {journal} {Phys. Rev. X}\ }\textbf {\bibinfo {volume} {5}},\ \bibinfo {pages}
  {031032} (\bibinfo {year} {2015})}\BibitemShut {NoStop}%
\bibitem [{\citenamefont {Potter}\ \emph {et~al.}(2015)\citenamefont {Potter},
  \citenamefont {Vasseur},\ and\ \citenamefont {Parameswaran}}]{PotterRSRG}%
  \BibitemOpen
  \bibfield  {author} {\bibinfo {author} {\bibfnamefont {A.~C.}\ \bibnamefont
  {Potter}}, \bibinfo {author} {\bibfnamefont {R.}~\bibnamefont {Vasseur}}, \
  and\ \bibinfo {author} {\bibfnamefont {S.~A.}\ \bibnamefont {Parameswaran}},\
  }\href {\doibase 10.1103/PhysRevX.5.031033} {\bibfield  {journal} {\bibinfo
  {journal} {Phys. Rev. X}\ }\textbf {\bibinfo {volume} {5}},\ \bibinfo {pages}
  {031033} (\bibinfo {year} {2015})}\BibitemShut {NoStop}%
\bibitem [{\citenamefont {Pal}\ and\ \citenamefont {Huse}(2010)}]{PalHuseMBL}%
  \BibitemOpen
  \bibfield  {author} {\bibinfo {author} {\bibfnamefont {A.}~\bibnamefont
  {Pal}}\ and\ \bibinfo {author} {\bibfnamefont {D.~A.}\ \bibnamefont {Huse}},\
  }\href {\doibase 10.1103/PhysRevB.82.174411} {\bibfield  {journal} {\bibinfo
  {journal} {Phys. Rev. B}\ }\textbf {\bibinfo {volume} {82}},\ \bibinfo
  {pages} {174411} (\bibinfo {year} {2010})}\BibitemShut {NoStop}%
\bibitem [{\citenamefont {Luitz}\ \emph {et~al.}(2015)\citenamefont {Luitz},
  \citenamefont {Laflorencie},\ and\ \citenamefont {Alet}}]{AletL22}%
  \BibitemOpen
  \bibfield  {author} {\bibinfo {author} {\bibfnamefont {D.~J.}\ \bibnamefont
  {Luitz}}, \bibinfo {author} {\bibfnamefont {N.}~\bibnamefont {Laflorencie}},
  \ and\ \bibinfo {author} {\bibfnamefont {F.}~\bibnamefont {Alet}},\ }\href
  {\doibase 10.1103/PhysRevB.91.081103} {\bibfield  {journal} {\bibinfo
  {journal} {Phys. Rev. B}\ }\textbf {\bibinfo {volume} {91}},\ \bibinfo
  {pages} {081103} (\bibinfo {year} {2015})}\BibitemShut {NoStop}%
\bibitem [{\citenamefont {Agarwal}\ \emph {et~al.}(2015)\citenamefont
  {Agarwal}, \citenamefont {Gopalakrishnan}, \citenamefont {Knap},
  \citenamefont {M\"uller},\ and\ \citenamefont {Demler}}]{AgarwalSubDiff}%
  \BibitemOpen
  \bibfield  {author} {\bibinfo {author} {\bibfnamefont {K.}~\bibnamefont
  {Agarwal}}, \bibinfo {author} {\bibfnamefont {S.}~\bibnamefont
  {Gopalakrishnan}}, \bibinfo {author} {\bibfnamefont {M.}~\bibnamefont
  {Knap}}, \bibinfo {author} {\bibfnamefont {M.}~\bibnamefont {M\"uller}}, \
  and\ \bibinfo {author} {\bibfnamefont {E.}~\bibnamefont {Demler}},\ }\href
  {\doibase 10.1103/PhysRevLett.114.160401} {\bibfield  {journal} {\bibinfo
  {journal} {Phys. Rev. Lett.}\ }\textbf {\bibinfo {volume} {114}},\ \bibinfo
  {pages} {160401} (\bibinfo {year} {2015})}\BibitemShut {NoStop}%
\bibitem [{\citenamefont {Bar~Lev}\ \emph {et~al.}(2015)\citenamefont
  {Bar~Lev}, \citenamefont {Cohen},\ and\ \citenamefont
  {Reichman}}]{BarLevSubDiff}%
  \BibitemOpen
  \bibfield  {author} {\bibinfo {author} {\bibfnamefont {Y.}~\bibnamefont
  {Bar~Lev}}, \bibinfo {author} {\bibfnamefont {G.}~\bibnamefont {Cohen}}, \
  and\ \bibinfo {author} {\bibfnamefont {D.~R.}\ \bibnamefont {Reichman}},\
  }\href {\doibase 10.1103/PhysRevLett.114.100601} {\bibfield  {journal}
  {\bibinfo  {journal} {Phys. Rev. Lett.}\ }\textbf {\bibinfo {volume} {114}},\
  \bibinfo {pages} {100601} (\bibinfo {year} {2015})}\BibitemShut {NoStop}%
\bibitem [{\citenamefont {{Lerose}}\ \emph {et~al.}(2015)\citenamefont
  {{Lerose}}, \citenamefont {{Kerala Varma}}, \citenamefont {{Pietracaprina}},
  \citenamefont {{Goold}},\ and\ \citenamefont
  {{Scardicchio}}}]{EnergyDiffusion}%
  \BibitemOpen
  \bibfield  {author} {\bibinfo {author} {\bibfnamefont {A.}~\bibnamefont
  {{Lerose}}}, \bibinfo {author} {\bibfnamefont {V.}~\bibnamefont {{Kerala
  Varma}}}, \bibinfo {author} {\bibfnamefont {F.}~\bibnamefont
  {{Pietracaprina}}}, \bibinfo {author} {\bibfnamefont {J.}~\bibnamefont
  {{Goold}}}, \ and\ \bibinfo {author} {\bibfnamefont {A.}~\bibnamefont
  {{Scardicchio}}},\ }\href {{http://arxiv.org/abs/1511.09144}} {\bibfield
  {journal} {\bibinfo  {journal} {ArXiv e-prints}\ } (\bibinfo {year}
  {2015})},\ \Eprint {http://arxiv.org/abs/1511.09144} {arXiv:1511.09144
  [cond-mat.dis-nn]} \BibitemShut {NoStop}%
\bibitem [{\citenamefont {{Bari{\v s}i{\'c}}}\ \emph
  {et~al.}(2016)\citenamefont {{Bari{\v s}i{\'c}}}, \citenamefont {{Kokalj}},
  \citenamefont {{Balog}},\ and\ \citenamefont {{Prelov{\v s}ek}}}]{Prelovsek}%
  \BibitemOpen
  \bibfield  {author} {\bibinfo {author} {\bibfnamefont {O.~S.}\ \bibnamefont
  {{Bari{\v s}i{\'c}}}}, \bibinfo {author} {\bibfnamefont {J.}~\bibnamefont
  {{Kokalj}}}, \bibinfo {author} {\bibfnamefont {I.}~\bibnamefont {{Balog}}}, \
  and\ \bibinfo {author} {\bibfnamefont {P.}~\bibnamefont {{Prelov{\v s}ek}}},\
  }\href@noop {} {\bibfield  {journal} {\bibinfo  {journal} {ArXiv e-prints}\ }
  (\bibinfo {year} {2016})},\ \Eprint {http://arxiv.org/abs/1603.01526}
  {arXiv:1603.01526 [cond-mat.str-el]} \BibitemShut {NoStop}%
\bibitem [{\citenamefont {Sirker}\ \emph {et~al.}(2009)\citenamefont {Sirker},
  \citenamefont {Pereira},\ and\ \citenamefont {Affleck}}]{affleckPRL}%
  \BibitemOpen
  \bibfield  {author} {\bibinfo {author} {\bibfnamefont {J.}~\bibnamefont
  {Sirker}}, \bibinfo {author} {\bibfnamefont {R.~G.}\ \bibnamefont {Pereira}},
  \ and\ \bibinfo {author} {\bibfnamefont {I.}~\bibnamefont {Affleck}},\ }\href
  {\doibase 10.1103/PhysRevLett.103.216602} {\bibfield  {journal} {\bibinfo
  {journal} {Phys. Rev. Lett.}\ }\textbf {\bibinfo {volume} {103}},\ \bibinfo
  {pages} {216602} (\bibinfo {year} {2009})}\BibitemShut {NoStop}%
\bibitem [{\citenamefont {Sirker}\ \emph {et~al.}(2011)\citenamefont {Sirker},
  \citenamefont {Pereira},\ and\ \citenamefont {Affleck}}]{AffleckHighTDiff}%
  \BibitemOpen
  \bibfield  {author} {\bibinfo {author} {\bibfnamefont {J.}~\bibnamefont
  {Sirker}}, \bibinfo {author} {\bibfnamefont {R.~G.}\ \bibnamefont {Pereira}},
  \ and\ \bibinfo {author} {\bibfnamefont {I.}~\bibnamefont {Affleck}},\ }\href
  {\doibase 10.1103/PhysRevB.83.035115} {\bibfield  {journal} {\bibinfo
  {journal} {Phys. Rev. B}\ }\textbf {\bibinfo {volume} {83}},\ \bibinfo
  {pages} {035115} (\bibinfo {year} {2011})}\BibitemShut {NoStop}%
\bibitem [{\citenamefont {B\"ohm}\ \emph {et~al.}(1994)\citenamefont {B\"ohm},
  \citenamefont {Viswanath}, \citenamefont {Stolze},\ and\ \citenamefont
  {M\"uller}}]{BhomMuller}%
  \BibitemOpen
  \bibfield  {author} {\bibinfo {author} {\bibfnamefont {M.}~\bibnamefont
  {B\"ohm}}, \bibinfo {author} {\bibfnamefont {V.~S.}\ \bibnamefont
  {Viswanath}}, \bibinfo {author} {\bibfnamefont {J.}~\bibnamefont {Stolze}}, \
  and\ \bibinfo {author} {\bibfnamefont {G.}~\bibnamefont {M\"uller}},\ }\href
  {\doibase 10.1103/PhysRevB.49.15669} {\bibfield  {journal} {\bibinfo
  {journal} {Phys. Rev. B}\ }\textbf {\bibinfo {volume} {49}},\ \bibinfo
  {pages} {15669} (\bibinfo {year} {1994})}\BibitemShut {NoStop}%
\bibitem [{\citenamefont {{Vasseur}}\ and\ \citenamefont
  {{Moore}}(2016)}]{JoelReview}%
  \BibitemOpen
  \bibfield  {author} {\bibinfo {author} {\bibfnamefont {R.}~\bibnamefont
  {{Vasseur}}}\ and\ \bibinfo {author} {\bibfnamefont {J.~E.}\ \bibnamefont
  {{Moore}}},\ }\href@noop {} {\bibfield  {journal} {\bibinfo  {journal} {ArXiv
  e-prints}\ } (\bibinfo {year} {2016})},\ \Eprint
  {http://arxiv.org/abs/1603.06618} {arXiv:1603.06618 [cond-mat.str-el]}
  \BibitemShut {NoStop}%
\bibitem [{\citenamefont {Steinigeweg}\ \emph {et~al.}(2015)\citenamefont
  {Steinigeweg}, \citenamefont {Gemmer},\ and\ \citenamefont
  {Brenig}}]{SteinigewegPRB}%
  \BibitemOpen
  \bibfield  {author} {\bibinfo {author} {\bibfnamefont {R.}~\bibnamefont
  {Steinigeweg}}, \bibinfo {author} {\bibfnamefont {J.}~\bibnamefont {Gemmer}},
  \ and\ \bibinfo {author} {\bibfnamefont {W.}~\bibnamefont {Brenig}},\ }\href
  {\doibase 10.1103/PhysRevB.91.104404} {\bibfield  {journal} {\bibinfo
  {journal} {Phys. Rev. B}\ }\textbf {\bibinfo {volume} {91}},\ \bibinfo
  {pages} {104404} (\bibinfo {year} {2015})}\BibitemShut {NoStop}%
\bibitem [{\citenamefont {Viswanath}\ and\ \citenamefont
  {Müller}(1994)}]{viswanath_recursion_book}%
  \BibitemOpen
  \bibfield  {author} {\bibinfo {author} {\bibfnamefont {V.~S.}\ \bibnamefont
  {Viswanath}}\ and\ \bibinfo {author} {\bibfnamefont {G.}~\bibnamefont
  {Müller}},\ }\href {http://www.springerlink.com/content/978-3-540-58319-6/}
  {\emph {\bibinfo {title} {The Recursion Method Application to Many-Body
  Dynamics}}},\ Lecture Notes in Physics monographs\ (\bibinfo  {publisher}
  {Springer Berlin Heidelberg},\ \bibinfo {year} {1994})\BibitemShut {NoStop}%
\bibitem [{\citenamefont {Lindner}\ and\ \citenamefont
  {Auerbach}(2010)}]{lindnerHCB}%
  \BibitemOpen
  \bibfield  {author} {\bibinfo {author} {\bibfnamefont {N.~H.}\ \bibnamefont
  {Lindner}}\ and\ \bibinfo {author} {\bibfnamefont {A.}~\bibnamefont
  {Auerbach}},\ }\href {\doibase 10.1103/PhysRevB.81.054512} {\bibfield
  {journal} {\bibinfo  {journal} {Phys. Rev. B}\ }\textbf {\bibinfo {volume}
  {81}},\ \bibinfo {pages} {054512} (\bibinfo {year} {2010})}\BibitemShut
  {NoStop}%
\bibitem [{\citenamefont {Fabricius}\ and\ \citenamefont
  {McCoy}(1998)}]{Fabricius1998}%
  \BibitemOpen
  \bibfield  {author} {\bibinfo {author} {\bibfnamefont {K.}~\bibnamefont
  {Fabricius}}\ and\ \bibinfo {author} {\bibfnamefont {B.~M.}\ \bibnamefont
  {McCoy}},\ }\href {\doibase 10.1103/PhysRevB.57.8340} {\bibfield  {journal}
  {\bibinfo  {journal} {Phys. Rev. B}\ }\textbf {\bibinfo {volume} {57}},\
  \bibinfo {pages} {8340} (\bibinfo {year} {1998})}\BibitemShut {NoStop}%
\bibitem [{\citenamefont {Coleman}(2015)}]{ColemanBook}%
  \BibitemOpen
  \bibfield  {author} {\bibinfo {author} {\bibfnamefont {P.}~\bibnamefont
  {Coleman}},\ }\href {http://dx.doi.org/10.1017/CBO9781139020916} {\emph
  {\bibinfo {title} {{Introduction to Many-Body Physics}}}}\ (\bibinfo
  {publisher} {Cambridge University Press},\ \bibinfo {year}
  {2015})\BibitemShut {NoStop}%
\bibitem [{com()}]{comment1}%
  \BibitemOpen
  \href@noop {} {}\bibinfo {note} {There are very few cases for which {$C(t)$}
  is known analytically. An exception is the quantum XY model (See Ref.
  35,36).}\BibitemShut {Stop}%
\bibitem [{\citenamefont {Perk}\ and\ \citenamefont {Capel}(1978)}]{Perk}%
  \BibitemOpen
  \bibfield  {author} {\bibinfo {author} {\bibfnamefont {J.}~\bibnamefont
  {Perk}}\ and\ \bibinfo {author} {\bibfnamefont {H.}~\bibnamefont {Capel}},\
  }\href {\doibase http://dx.doi.org/10.1016/0378-4371(78)90026-2} {\bibfield
  {journal} {\bibinfo  {journal} {Physica A: Statistical Mechanics and its
  Applications}\ }\textbf {\bibinfo {volume} {92}},\ \bibinfo {pages} {163 }
  (\bibinfo {year} {1978})}\BibitemShut {NoStop}%
\bibitem [{\citenamefont {Roldan}\ \emph {et~al.}(1986)\citenamefont {Roldan},
  \citenamefont {McCoy},\ and\ \citenamefont {Perk}}]{McCoy}%
  \BibitemOpen
  \bibfield  {author} {\bibinfo {author} {\bibfnamefont {J.}~\bibnamefont
  {Roldan}}, \bibinfo {author} {\bibfnamefont {B.~M.}\ \bibnamefont {McCoy}}, \
  and\ \bibinfo {author} {\bibfnamefont {J.~H.}\ \bibnamefont {Perk}},\ }\href
  {\doibase http://dx.doi.org/10.1016/0378-4371(86)90254-2} {\bibfield
  {journal} {\bibinfo  {journal} {Physica A: Statistical Mechanics and its
  Applications}\ }\textbf {\bibinfo {volume} {136}},\ \bibinfo {pages} {255 }
  (\bibinfo {year} {1986})}\BibitemShut {NoStop}%
\bibitem [{\citenamefont {Mori}(1965)}]{Mori1965}%
  \BibitemOpen
  \bibfield  {author} {\bibinfo {author} {\bibfnamefont {H.}~\bibnamefont
  {Mori}},\ }\href {\doibase 10.1143/PTP.34.399} {\bibfield  {journal}
  {\bibinfo  {journal} {Progress of Theoretical Physics}\ }\textbf {\bibinfo
  {volume} {34}},\ \bibinfo {pages} {399} (\bibinfo {year} {1965})}\BibitemShut
  {NoStop}%
\bibitem [{\citenamefont {Lee}(1982)}]{Lee1982}%
  \BibitemOpen
  \bibfield  {author} {\bibinfo {author} {\bibfnamefont {M.~H.}\ \bibnamefont
  {Lee}},\ }\href {\doibase 10.1103/PhysRevB.26.2547} {\bibfield  {journal}
  {\bibinfo  {journal} {Phys. Rev. B}\ }\textbf {\bibinfo {volume} {26}},\
  \bibinfo {pages} {2547} (\bibinfo {year} {1982})}\BibitemShut {NoStop}%
\end{thebibliography}
%
\end{document}